\documentclass[twocolumn,prb]{revtex4}% Physical Review B
\usepackage{graphicx}% Include figure files
\usepackage{dcolumn}% Align table columns on decimal point
\usepackage{bm}% bold math
\usepackage{amssymb}%Use amssymb package
\newcommand{\nic}{Ni$^{2+}$}
\newcommand{\mg}{Mg$^{2+}$}
\newcommand{\zn}{Zn$^{2+}$}

\newcommand{\cu}{Cu$^{2+}$}

\newcommand{\cgo}{CuGeO$_3$}
\newcommand{\cugeo}{CuGeO$_3$}
\newcommand{\pbn}{PbNi$_2$V$_2$O$_8$}

\newcommand{\fig}[2]{\begin{center}\includegraphics[width=#1cm,clip,keepaspectratio]{#2}\end{center}}
\newcommand{\pbnmn}{Pb(Ni$_{1-x}$Mn$_x$)$_2$V$_2$O$_8$}
\newcommand{\pbnmg}{Pb(Ni$_{1-x}$Mg$_x$)$_2$V$_2$O$_8$}
\newcommand{\pbnco}{Pb(Ni$_{1-x}$Co$_x$)$_2$V$_2$O$_8$}
\newcommand{\pbncu}{Pb(Ni$_{1-x}$Cu$_x$)$_2$V$_2$O$_8$}
\newcommand{\pbnm}{Pb(Ni$_{1-x}M_x$)$_2$V$_2$O$_8$}

\date{\today}

\begin{document}

\title{Ordered State in a Haldane Material PbNi$_2$V$_2$O$_8$ Doped
with Magnetic and Non-Magnetic Impurities}

\author{S.\ Imai}
 \altaffiliation{Electronic address: imai.suguru@furukawa.co.jp.
 Present address: The Furukawa Electric Co., Ltd. 2-4-3 Okano, Nishi-ku, Yokohama
 220-0073, Japan
}
 %\affiliation{Department of Advanced Materials Science, the University of Tokyo, }%Lines break automatically or can be forced with \\
\author{T.\ Masuda}
 \altaffiliation{Electronic address: masudat@ornl.gov Present address: Condensed Matter Sciences Division, Oak Ridge
 National Laboratory,
 Oak Ridge, Tennessee 37831-6393, USA}
\author{T.\ Matsuoka}
 \altaffiliation{Present address: System Infrastructure Technologies, System Research \& Development Center,
 NS Solutions Corporation, 3-3-1, Minatomirai, Nishi-ku, Yokohama 220-8401, Japan }
\author{K.\ Uchinokura}
 \altaffiliation{Present address: The Institute of Physical and Chemical
  Research (RIKEN), Wako, Saitama 351-0198, Japan}
\affiliation{Department of Advanced Materials Science, The University of Tokyo,
5-1-5 Kashiwa-no-ha, Kashiwa 277-8581, Japan}
%\address{}

%%%%%% ABSTRACT %%%%%% ABSTRACT %%%%%% ABSTRACT %%%%%% ABSTRACT %%%%%%
\begin{abstract}
Impurity effect is systematically studied in doped Haldane
material \pbnm\ ($M$ = Mn, Co, Cu, and Mg) by use of DC and AC
susceptibility, and heat capacity measurements. The occurrence of
three-dimensional ordered state is universally observed for all
the impurities and the complete temperature -- concentration phase
diagrams are obtained, which are qualitatively similar to that in
other spin-gap materials. The unique feature is found in the
drastic dependence of the transition temperatures on the species
of the impurities. The consideration of effective Hamiltonian
based on VBS model makes it clear that the ferromagnetic
next-nearest-neighbor interaction and the antiferromagnetic
nearest-neighbor interaction between impurity and edge spins play
a key role in the unique feature.
\end{abstract}

\pacs{75.10.Jm, 75.30.Kz, 75.50.Ee}

\maketitle

%%%%%% INTRODUCTION %%%%%% INTRODUCTION %%%%%% INTRODUCTION %%%%%%
\section{Introduction\label{I}}
\begin{figure}[h!]
    \fig{8}{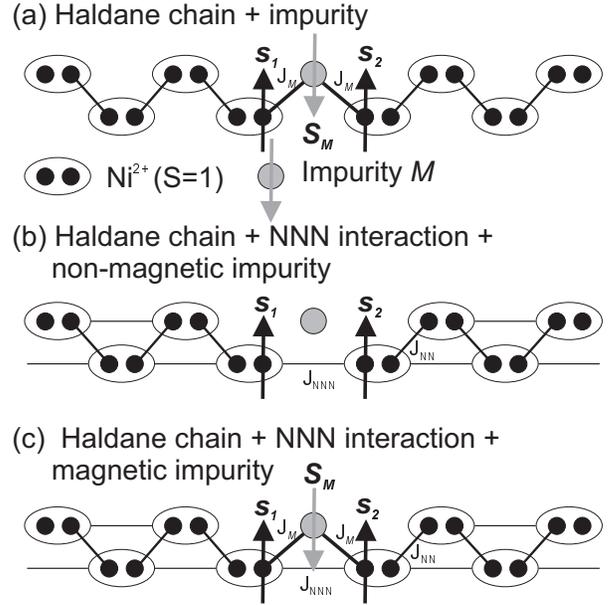}
\caption{(a) Effective spin configuration in doped-Haldane chain. $S$ = 1
spins are described by a couple of $S$ = 1/2 spins (filled
circles), and symmetrization operator (ellipsoids). Adjacent $S$ =
1/2 spins on different Ni$^{2+}$ sites are coupled by singlet bond
(solid line). (b) Effective spin
configuration in non-magnetic impurity doped Haldane chain with
next-nearest-neighbor interaction. The next-nearest-neighbor
 interaction $J_{\rm NNN}$ and nearest-neighbor interaction $J_{\rm NN}$
 is indicated by thin and bold solid lines, respectively.
(c) Effective spin configuration in magnetic impurity doped
Haldane chain with next-nearest-neighbor interaction. }
\label{fig:spins}
\end{figure}

Low-dimensional quantum magnetism is one of the most exciting
topics in condensed matter physics. The simplest model is realized
in one-dimensional Heisenberg antiferromagnet, where the
Hamiltonian is expressed as
\begin{equation}
\mathcal{H}=J\sum_{j}\bm{S}_{j}\cdot\bm{S}_{j+1}, \qquad J>0.
\label{Hamiltonian}
\end{equation}
Exact ground state in $S=1/2$ is non-magnetic
singlet~\cite{31bethe} and its excitation is gapless known as des
Cloizeaux-Pearson mode.\cite{62desCloizeaux} In 1983 Haldane
pointed out a qualitative difference between spin integer and half
integer in Eq.~(\ref{Hamiltonian}); in the former spin correlation
decays exponentially and spin excitation has a spin gap, while in
the latter the correlation decays by power law and the excitation
is gapless.\cite{83haldane1,83haldane2}
%Since then the Haldane gap
%has been observed in several quasi-one-dimensional
%antiferromagnets\cite{87renard,88renard,93darriet} and the Haldane
%chain is one of the current interest in quantum magnetism as well
%as other spin gap models.\cite{75bray,79cross,93hase,92dagotto,97garrett}
%
A mimic model of spin integer Heisenberg antiferromagnetic chain
known as AKLT (Affleck, Kennedy, Lieb, and Tasaki)
Hamiltonian\cite{87affleck,88affleck} was proposed by adding a
quadratic term in Eq.~(\ref{Hamiltonian}). AKLT Hamiltonian can be
solved analytically and the ground state, which is known as
valence-bond-solid (VBS) state, is similar to that of simple
Heisenberg Hamiltonian. In a finite chain staggered spins are
induced in the vicinity of its edges\cite{87affleck,88affleck} and
they can be approximated as a couple of effective $S = 1/2$ spins
on edges. Edge spins predicted by AKLT Hamiltonian is realized in
the impurity-doped Haldane chain, as is schematized in
Fig.~\ref{fig:spins} (a), and the low energy excitation can be
expressed by the effective Hamiltonian consisting of three spins
with open boundary condition;
\begin{equation}
\mathcal{H}=J_M(\bm{s}_1\cdot\bm{S}_{M}+\bm{S}_{M}\cdot\bm{s}_2).
\label{effective_Hamiltonian}
\end{equation}
The predicted energy excitations were observed in the ESR spectrum
in impurity-doped Haldane material
NENP~(Ref.~\onlinecite{90hagiwara}) and numerical calculation also
supports application of AKLT Hamiltonian to Haldane
chain.\cite{90kennedy,93miyashita}

%%%%%%%%%%%%%%%%%%%%%%%%%%%%%%%%%%%%%%%%%%%%%%%%%%

\pbn\ is a new member of inorganic Haldane materials discovered by
Uchiyama \textit{et~al.} in 1999.\cite{99uchiyama} The Haldane gap
was confirmed by the bulk susceptibility, high-field
magnetization, and inelastic neutron
scattering.\cite{99uchiyama,00zheludev} The unique feature in
\pbn\ is spin-vacancy-induced antiferromagnetic phase in \pbnmg
,\cite{99uchiyama,00uchinokura} where coexistence of gapless and
gap excitations is suggested. The occurrence of the
antiferromagnetic phase is attributed to the {\it ferromagnetic
next-nearest-neighbor interaction} and the {\it interchain
interaction}. The former is indicated by powder neutron
scattering\cite{01zheludev} and ESR\cite{02zorko} measurements and
decisively confirmed by the consideration of Schottky heat
capacity.\cite{02masuda} The latter is confirmed by careful
analysis of powder neutron scattering
measurements.\cite{00zheludev} In the antiferromagnetic
long-range-ordered state, spins adjacent to impurity must be
parallel as is schematized in Fig.~\ref{fig:spins} (b). Therefore,
ferromagnetic next-nearest-neighbor interaction favors
antiferromagnetic ordering. Interchain interaction is, of course,
necessary for the three-dimensional ordering due to Mermin-Wagner
theorem.\cite{66mermin}

Since the low energy excitation in doped NENP was well described
by the effective Hamiltonian Eq.~(\ref{effective_Hamiltonian}), a
similar approach could be useful to understand the ordered state
of doped \pbn . Effective Hamiltonian for Mg$^{2+}$ ($S$ =
0)-doped \pbn\ will be
\begin{equation}
\mathcal{H}_1=J_{\rm NNN}\bm{s}_1\cdot\bm{s}_2, \qquad J_{\rm
NNN}<0, \label{effective_Mg}
\end{equation}
which is actually the Hamiltonian for ferromagnetic dimer. The
ground state is $S$ = 1 triplet states, where the spins are
parallel to each other, and favors three-dimensional ordering. On
the other hand the effective Hamiltonian for \pbn\ doped with a
magnetic impurity is expressed as three spin ring as is shown in
Fig.~\ref{fig:spins} (c);
\begin{equation}
\mathcal{H}_2=J_M(\bm{s}_1\cdot\bm{S}_{M}+\bm{S}_{M}\cdot\bm{s}_2)+J_{\rm
NNN}\bm{s}_1\cdot\bm{s}_2. \label{effective_M}
\end{equation}
In this case the ground state depends on the sign of $J_M$ and
also the type of spin interaction such as Heisenberg, Ising, and
XY. Because they are subject to the species of impurity ions $M$,
systematic study on doped \pbn\ may be valuable to reveal the
nature of the interaction among the impurity spins and edge spins.
Although there have been several studies in non-magnetic
impurity-doped \pbn
,\cite{99uchiyama,00uchinokura,02zorko,02masuda,02smirnov} study
of magnetic-impurity doped \pbn\ is very rare and even
preliminary.\cite{01uchinokura} We will report a complete study in
\pbnm\ ($M$ = Mn, Co, Cu, and Mg) and have found the drastic
dependence on the kind of impurities in the temperature --
concentration ($T$--$x$) phase diagram.

To study the magnetic excitations neutron inelastic scattering would be
the most effective tool if a single crystal of reasonable size were
available. However, \pbn\ is incongruent and it is hard to
obtain large single crystal. Therefore the best we can do is bulk
measurements such as magnetic susceptibility and heat capacity on
polycrystalline samples. Experimental details will be briefly
mentioned in section II. In section III A and C the value and
number of effective spins induced by impurities are carefully
discussed and the sign of $J_M$ are obtained. The value of
effective spin is estimated to be $|S_M-1|$ in \pbnm\ ($M$ = Mn,
Co, and Cu) and $J_M$ is proved to be antiferromagnetic. In
section III B antiferromagntic transition is confirmed in \pbncu\
at extremely low temperatures. In section III D the $T$--$x$ phase
diagrams are obtained in \pbnm\ ($M$ = Mn, Co, Mg, and Cu). In
section IV the weak ferromagnetism in \pbnco , the drastic
dependence of the transition temperatures on the species of the
impurities in $T$--$x$ phase diagram, and the comparison with
impurity-doped spin-Peierls material \cugeo\ are discussed.

%%%%%%%% EXPERIMENTAL DETAILS %%%%%%%%%% EXPERIMENTAL DETAILS %%%%%%%%%%
\section{Experimental Details}
Polycrystalline samples of \pbnm\ ($M$ = Mn, Co, Mg, and Cu) were
prepared by solid state reaction method. Aligned samples were
prepared by applying magnetic field to the mixture of powder
sample and epoxy resin (Stycast 1266) at room
temperature.\cite{99uchiyama} DC magnetic susceptibilities were
measured down to 2.0~K in $H=0.1$~T magnetic field by SQUID
magnetometer (MPMS-XL of Quantum Design Corp.) AC susceptibilities
were measured down to 50~mK in zero static magnetic field by using
an adiabatic demagnetization refrigerator ($\mu$Fridge of
Cambridge Magnetic Refrigeration Corp.). Heat capacities were
measured down to 0.4~K by the relaxation method (PPMS with
Helium-3 option of Quantum Design Corp.) in a magnetic field up to
12~T.

%%%%%% EXPERIMENT %%%%%% EXPERIMENT %%%%%% EXPERIMENT %%%%%%
\section{Experimental Results}
\subsection{DC Magnetic Susceptibility\label{IIIA}}

\begin{figure}[h!]
    \fig{8}{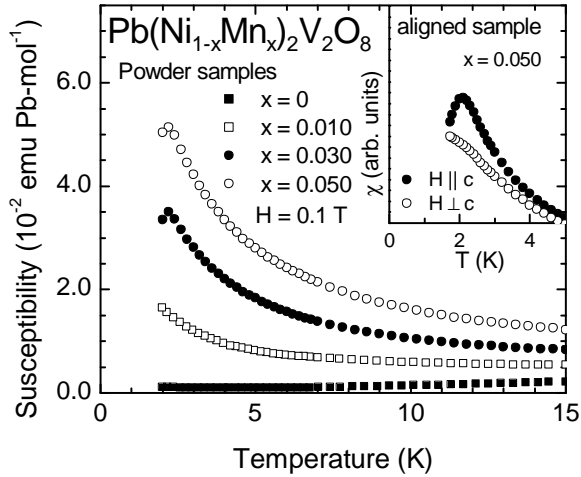}
\caption{Temperature dependence of the magnetic susceptibilities
in \pbnmn. Inset shows the anisotropy of the aligned
polycrystalline sample with $x=0.050$. } \label{fig:pbnmn-mt}
\end{figure}

Figure~\ref{fig:pbnmn-mt} shows that a magnetic anomaly is
detected in the DC magnetic susceptibility in \pbnmn\ ($x=0.050$)
aligned sample. A sharp peak in $\bm{H}
\parallel c$ axis and a slight anomaly in $\bm{H} \perp c$ at 2.3
K in the inset of Fig.~\ref{fig:pbnmn-mt} confirms that the ground
state is an easy-axis type antiferromagnetic long-range order. The
phase transition is observed also in the polycrystalline sample of
$x=0.030$. No anomaly in the lower concentration
sample $x=0.010$ is simply because of our experimental limit.

%%%%%%%%%%%%%%%%%%%%%%%%%%%%%%%%%%%%%%%%%%%%%%%%%%
%%%%%%%%%%%%%%%%%%%%%%%%%%%%%%%%%%%%%%%%%%%%%%%%%%55
\begin{figure}[h!]
    \fig{8}{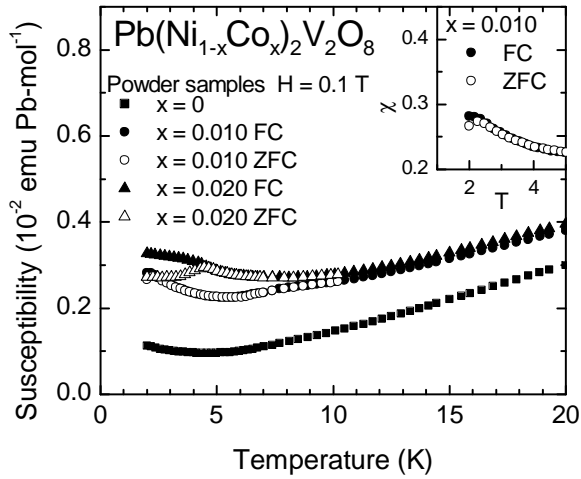}
\caption{Temperature dependence of the magnetic susceptibilities
in \pbnco . The inset is the low temperature susceptibility in $x$
= 0.01 sample. } \label{fig:pbnco-mt}
\end{figure}
%%%%%%%%%%%%%%%%%%%%%%%%%%%%%%%%%%%%%%%%%%%%%%%%%%%%%%%
\begin{figure}[h!]
    \begin{center}
        \includegraphics*[width=8cm,clip]{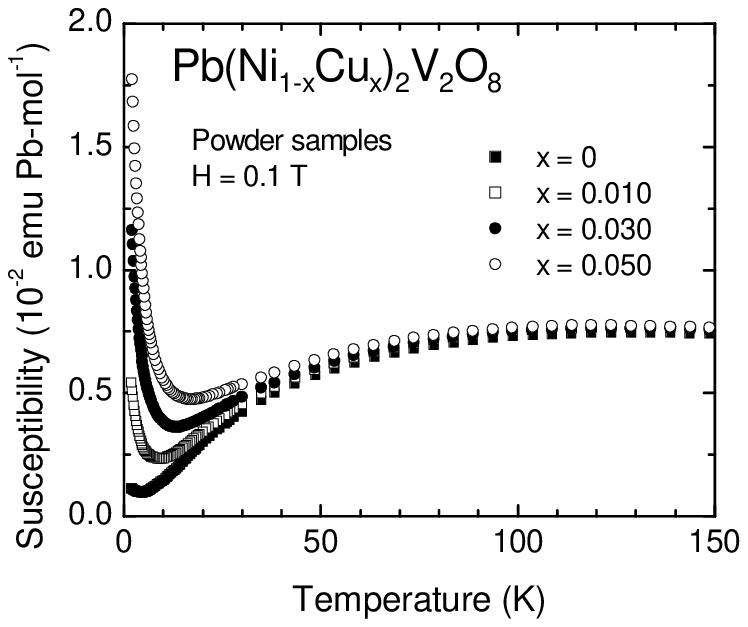}
    \end{center}
    \caption{Temperature dependence of the magnetic susceptibilities
    in
    \pbncu.}
    %\fig{8}{pbncu-mt.eps}
\label{fig:pbncu-mt}
\end{figure}
%%%%%%%%%%%%%%%%%%%%%%%%%%%%%%%%%%%%%%%%%%%%%%%%%%%

\begin{figure}[h!]
    \fig{8}{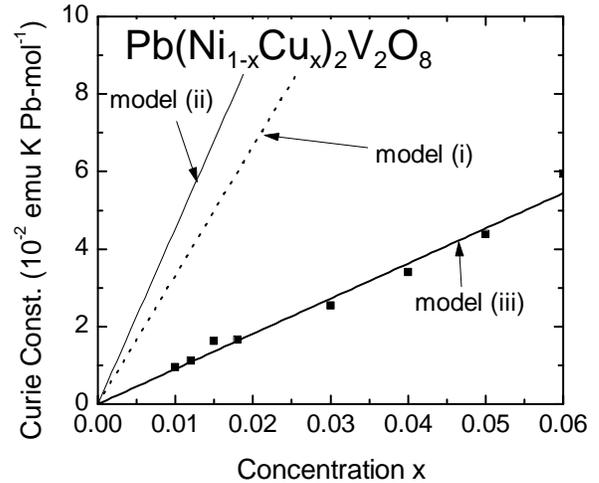}
\caption{Curie constants of \pbncu\ from the measurement of the
magnetic susceptibility and the calculated values from the three
models.} \label{fig:pbncu-curie}
\end{figure}

%%%%%%%%%%%%%%%%%%%%%%%%%%%%%%%%%%%%%%%%%%%%%%%%%%
\begin{figure}[h!]
    \fig{8}{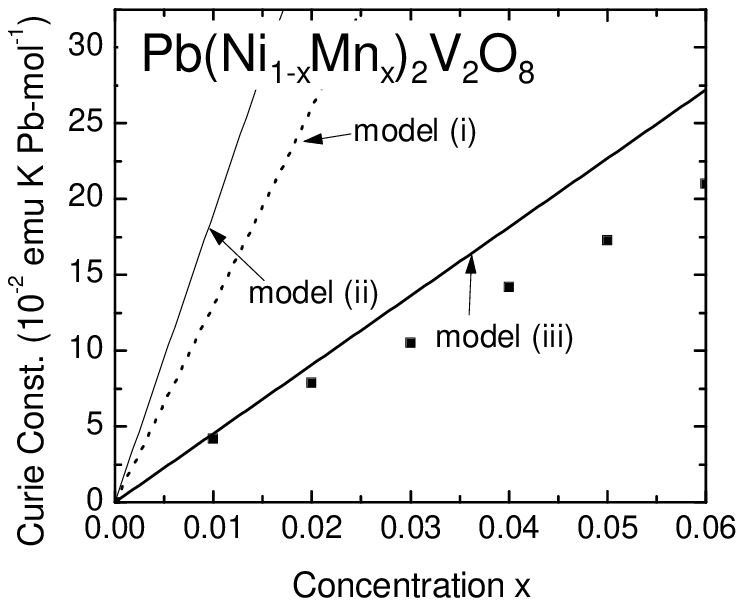}
\caption{Curie constants of \pbnmn\ from the measurement of the
magnetic susceptibility and the calculated values from the three
models. } \label{fig:pbnmn-curie}
\end{figure}
%%%%%%%%%%%%%%%%%%%%%%%%%%%%%%%%%%%%%%%%%%%%%%%%%%

%%%%%%%%%%%%%%%%%%%%%%%%%%%%%%%%%%%%%%%%%%%%%%%%%%

Figure~\ref{fig:pbnco-mt} shows the temperature dependence of the
magnetic susceptibilities in \pbnco. They were measured both in a
field cooling (FC) and in a zero-field cooling (ZFC) processes.
The impurity-induced anomalies are observed at $T=2.2$~K for
$x=0.010$ and $T=4.2$~K for $x=0.020$, and the hysteresis between
FC and ZFC were detected below the anomaly's temperature. The
magnetic susceptibilities seem to have a finite values for both
processes at $T = 0$ K, which would be attributed to an
antiferromagnetic transition with weak spontaneous magnetization.
In mixed spin materials spin glass is another possibility but the
jump in the heat capacity, which we will see later, prefers weak
ferromagnetism.  The weak ferromagnetism is possible owing to the
existence of the Dzyaloshinskii-Moriya antisymmetric
interaction\cite{57dzyalo,60moriya} as will be discussed in the
section \ref{IV}.

No anomalies are detected in any concentration samples of \pbncu\
in the measurements down to $T$ = 2 K. In contrast with other
impurities Cu-doping only enhances Curie-like susceptibility at
low temperatures which is shown in Fig.~\ref{fig:pbncu-mt}.

Hereafter we will estimate the number and the value of
impurity-induced effective spins based on VBS model, where the
magnetization in a finite chain is presumed to be induced only by
the edge spins.\cite{87affleck,88affleck} Let us assume that  the
susceptibility is expressed as a sum of the two contributions in
the low impurity concentration region:
%%%%%%%%%%%%%%%%%%%%%%%%%%%%%%%%%%%%%%
\begin{equation}
\chi=\chi_{\mathrm{pure}}+\chi_{\mathrm{para}},
\end{equation}
%%%%%%%%%%%%%%%%%%%%%%%%%%%%%%%%%%%%%%%
where $\chi_{\mathrm{pure}}$ is the susceptibility in pure \pbn\
and $\chi_{\mathrm{para}}$ is the additional susceptibility due to
impurity-induced spins. If $\chi_{\mathrm{para}}$ obeys
Curie-Weiss law, the number and the value of the
impurity-induced-effective spins will be obtained from Curie
constants. As was already mentioned in the introduction the {\it
ferromagnetic next-nearest-neighbor interaction} exists, which is
in fact confirmed by the observation of effective spin with $S$ =
1 in \pbnmg\ by some of the present authors.\cite{02masuda}
Therefore, the following three models are possible in
impurity-doped \pbn ;
%%%%%%%%%%%%%%%%%%%%%%%%%%%%%%%%%%%%%%%%%%%%
\begin{enumerate}
\item[(i)] The interaction between the edge spin and the impurity
spin, $J_M$ in Eq. (\ref{effective_M}), is very small and edge
spins are coupled by ferromagnetic next-nearest-neighbor
interaction. One effective spin with $S$ = 1 and one with $S =
S_M$ will be detected.
\item[(ii)]  $J_M$ is ferromagnetic and
one effective spin with $S=S_M+1$ will be detected.
\item[(iii)] $J_M$ is antiferromagnetic and
one effective spin with $S=|S_M-1|$ will be detected.
%\item[(iv)] The interaction between the edge spin and the impurity
%spin is very weak and the interaction between edge spins of a
%segment in a chain is strong. The effective spin of the chain with
%even \nic\ ions is zero and that of the chain with odd \nic\ ions
%is one. This is called ``singlet-triplet model" which is observed
%only in the doped Haldane material
%Y$_2$BaNi$_{1-x}$Zn$_x$O$_5$.\cite{94ramirez}
\end{enumerate}
%%%%%%%%%%%%%%%%%%%%%%%%%%%%%%%%%%%%%%%%%%

Curie constants of \pbncu\ are plotted as a function of $x$ in
Fig.~\ref{fig:pbncu-curie}. Three lines are calculated values from
the above three models.
%\question{under the assumption of $g=2.2$. See the above question.}
The experimental data are close to the theoretical line of
the model (iii) and
this suggests that  $J_{\rm Cu}$
in Eq. (\ref{effective_M}) is antiferromagnetic.

In the same way $\chi_{\mathrm{para}}$ of \pbnmn\ were fitted by
the Curie-Weiss law and the experimental Curie constants and the
calculated values of the three models are plotted in
Fig.~\ref{fig:pbnmn-curie}. The same argument on \pbncu\ may lead
to the conclusion that the most suitable theoretical value is the
model (iii),  at least for $x\rightarrow 0$. Hence the
impurity-induced spins have $S=3/2$ degree of freedom at each
impurity site and $J_{\rm Mn}$ is antiferromagnetic.

We assumed $g$ = 2.2 implicitly because Zorko
\textit{et~al.}\cite{02zorko} has shown that the broad ESR
resonance has $g\sim 2.2$ between room temperature and 70~K for
\pbnmg\ ($x\le 0.24$). However, we have to be careful about it
because the ESR resonance was attributed to the bulk spins (on
\nic\ ions) and not that on impurity ions. At low temperatures
they observed the shift of the resonance line, which can be due to
the growing contribution of impurity-related
resonance.\cite{02zorko} Therefore it might be difficult to
definitely determine which of the three models is the best
description of the spin state only by the susceptibility
measurement. Since magnetic entropy also gives information on the
value and the number of spins and it is independent on $g$ value,
heat-capacity measurements might be more useful to discuss the
model precisely. We will see that our argument here is consistent
with heat capacity measurements in III C.

%%%%%%%%%%%%%%%%%%%%%%%%%%%%%%%%%%%%%
\subsection{AC Susceptibility at Very Low Temperatures\label{IIIB}}
\begin{figure}[h!]
    \fig{8}{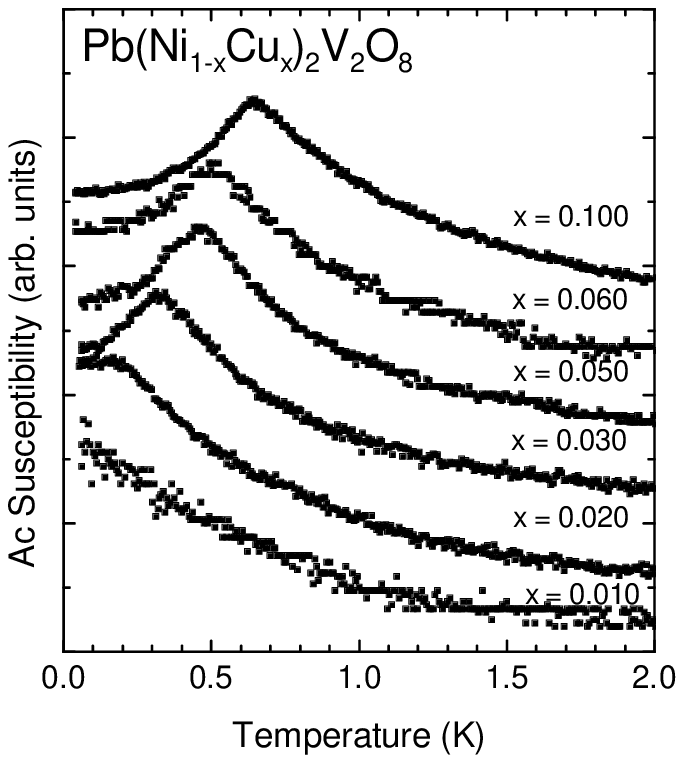}
\caption{Temperature dependence of the AC susceptibilities in
\pbncu\ at very low temperatures.} \label{fig:pbncu-mtl}
\end{figure}

%%%%%%%%%%%%%%%%%%%%%%%%%%%%%%%%%%%%%
Although the DC magnetic susceptibility and heat capacity in
\pbncu\ have been measured down to 2 K in the same way as the
samples doped with other species of impurities, three-dimensional
ordering has not been found yet\cite{01uchinokura} (see also
\ref{IIIA}). There arizes a question if there is no transition in 
\pbncu\ or the transition temperature is extremely low. 
% (less than $\lesssim0.5$~K).
To search for the magnetic phase transition in \pbncu,
measurements of AC susceptibility at low temperatures (down to
50~mK) were performed using the adiabatic demagnetization method.
The real parts of the AC susceptibilities in \pbncu\ at very low
temperatures are shown in Fig.~\ref{fig:pbncu-mtl}; here the
absolute value was not exactly obtained because of the
specification of the AC susceptibility method but it is sufficient
to determine the existence or absence of the transition and also
to determine the transition temperature if it exists. The external
static magnetic field was not applied in this experiment.

For the sample with $x=0.010$, the change of the susceptibility is
likely to follow the Curie-Weiss law and no anomaly was found
above 50~mK. For the sample with $x=0.020$, however, an anomaly
appears in the susceptibility-temperature curve at $\sim 0.2$~K
and the susceptibility-temperature curve becomes flat below this
temperature. For the samples with $x\ge 0.30$ we observed a
distinct peak in the susceptibility-temperature curve and the peak
temperature increases with the impurity (\cu) concentration.
%\question{No hysteresis was found in AC susceptibility-temperature curves
%in any of the measured samples. Correct?}

This behavior is qualitatively the same as that of \pbnmg , which
has three-dimensional ordering at low temperatures, and we
attributed the peaks in \pbncu\ to magnetic orderings. Hence, we
concluded that the ground state of \pbncu\ is also the ordered
state as the samples doped with other impurities. However, temperature
scale of the ordering temperature is quite different. This will be
discussed in section IV.

%%%%%%%%%%%%%%%%%%%%%%%%%%%%%%%%%%%%%%%
\subsection{Heat Capacity\label{IIIC}}
\begin{figure}[h!]
    \fig{8}{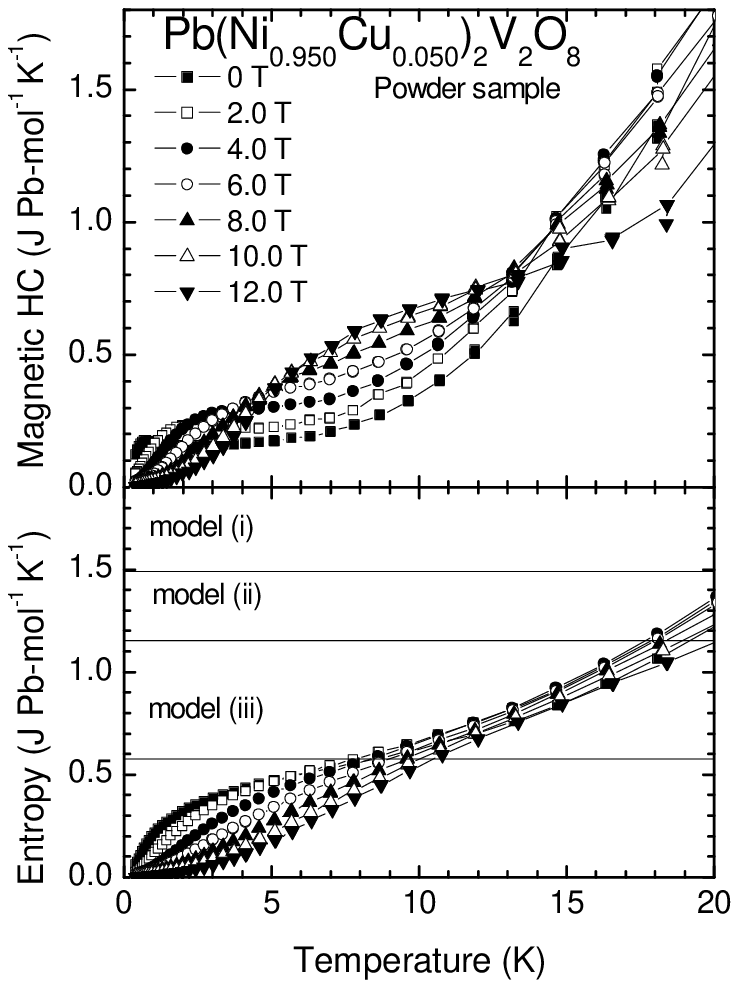}
\caption{Magnetic heat capacity (upper panel) and magnetic entropy
(lower panel) of \pbncu\ with $x=0.050$.}
\label{fig:pbncu-hc}
\end{figure}

%%%%%%%%%%%%%%%%%%%%%%%%%%%%%%%%%%%%%%%
We obtained the magnetic heat capacity $C_{\mathrm{m}}$ by
subtracting the heat capacity of PbMg$_2$V$_2$O$_8$, which is
isostructural to \pbn\, from that of \pbnm . The magnetic entropy
$S_{\mathrm{m}}$ was obtained according to the equation
$S_{\mathrm{m}}=\int_{0}^{T} \frac{C_{\mathrm{m}}}{T}dT$. Figure
\ref{fig:pbncu-hc} shows the magnetic heat capacity (upper panel)
and magnetic entropy (lower panel) of \pbncu\ with $x=0.050$. For
$H=0$~T a small anomaly is observed at $\sim 0.5$~K, which is
probably attributed to the antiferromagnetic ordering because the
temperature is close to the peak temperature, 0.46~K, in the AC
susceptibility. This anomaly diminishes by the application of a
magnetic field. This is the typical behavior of classical
antiferromagnetic transition. In magnetic field there appears a
Schottky-like broad anomaly at higher temperature and the
anomaly's temperature increases with the magnetic field. This
behavior was already reported in \pbnmg\
(Ref.~\onlinecite{02masuda}) and the difference is the energy
scale of the ordered state such as the transition temperature or
the critical field for the disappearance of the ordered state.

%%%%%%%%%%%%%%%%%%%%%%%%%%%%%%%%%%%%%%%%%%%%%%%%%%%%%%%
%%%%%%%%%%%%%%%%%%%%%%%%%%%%%%%%%%%%%%%%%%%%%%%%%%%%%%%
\begin{figure}[h!]
    \fig{8}{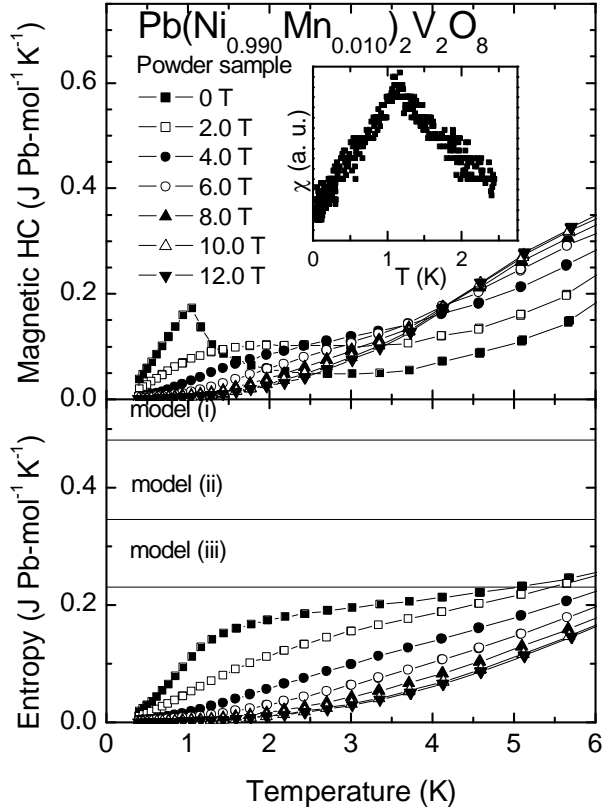}
\caption{Magnetic heat capacity (upper panel) and magnetic entropy
(lower panel) of \pbnmn\ with $x=0.010$. The inset shows AC
susceptibility in the same sample measured in the adiabatic
demagnetization refrigerator. } \label{fig:pbnmn-hc}
\end{figure}
%%%%%%%%%%%%%%%%%%%%%%%%%%%%%%%%%%%%%%%%%%%%%%%%%%%%%%%
\begin{figure}[h!]
    \fig{8}{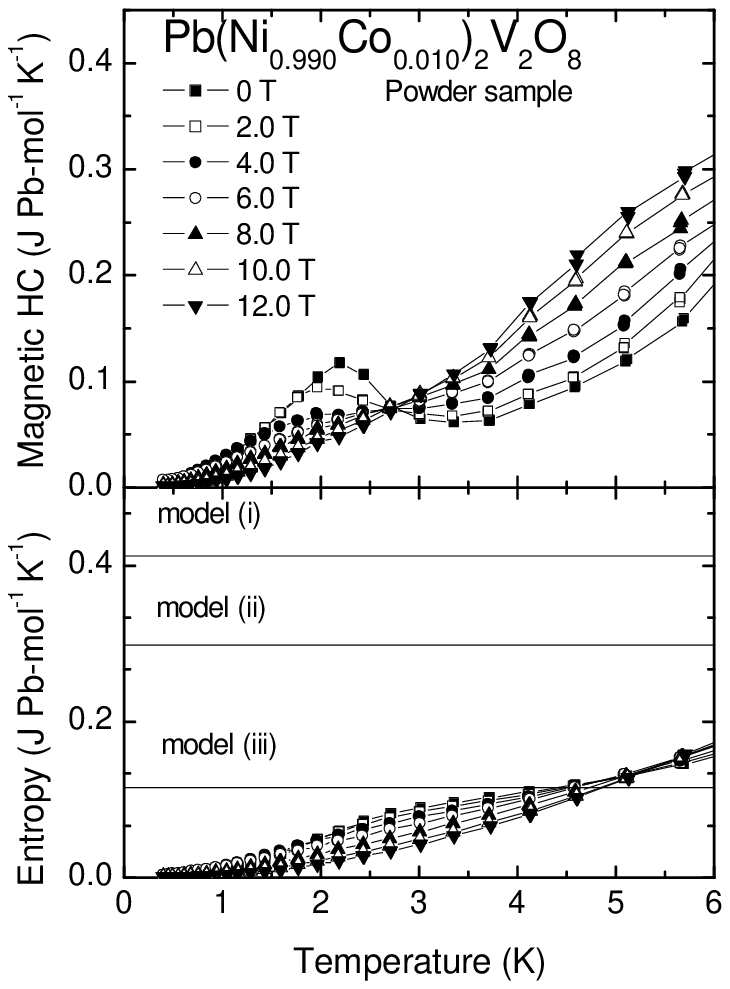}
\caption{Magnetic heat capacity (upper panel) and magnetic entropy
(lower panel) of \pbnco\ with $x=0.010$.
}
\label{fig:pbnco-hc}
\end{figure}
%%%%%%%%%%%%%%%%%%%%%%%%%%%%%%%%%%%%%%%%%%%%%%%%%%%%%%%

If we assume that the system can be approximated by an assembly of
the free spins (effective spins) of the same $S$ value, the
entropy of the system will be saturated at
$S_{\mathrm{m}}=Nk_{\mathrm{B}}\ln (2S+1)$. Note that the above
expression does not contain the $g$ value. Therefore the magnetic
entropy would be more appropriate quantity to obtain the value and
the number of spins than the Curie constant. The saturation values
of the three models which is described in section III A are shown
in the lower panel of Fig.~\ref{fig:pbncu-hc} (solid lines). We
can see that the entropies at the various values of the magnetic
fields converge to almost the same values at $T \sim 12$ K and a
point of inflection seems to exist there.

In addition to the induced moments as effective spins, there exist
other contributions to magnetic entropy.
\begin{enumerate}
\item[(a)] One is the contribution from the spin-gap excitation.
Haldane-like gap excitation, in which the energy scale is about 20
K, was observed in \pbnmg\ by the high field magnetization
measurement\cite{02masuda} and coexistence of gapless and gap
excitations, which was originally discovered in doped spin-Peierls
material \cugeo , was suggested in the ordered phase. Although the
contribution from the Haldane-like gap excitation is small at low
temperatures, it increases with temperature.
\item[(b)] Another possible source of the error may come
from the subtraction of lattice heat capacity; the assumption that
heat capacity in PbMg$_2$V$_2$O$_8$ is lattice heat capacity in
\pbnm\ might not be perfect.
\end{enumerate}
Compared with the contribution of (a), that of (b) would be
negligible in the temperature regions of Figs.~\ref{fig:pbncu-hc},
because the lattice heat capacity changes as $\sim T^3$.
%
%In the lower panel of Fig.~\ref{fig:pbncu-hc}, magnetic entropy at
%$H =$ 0 T increases rapidly up to the transition temperature while
%the entropy slowly increases above the transition temperature. The
%magnetic entropies in various magnetic fields converge to almost
%the same value at about 12~K and a point of curvature seems to
%exist there. This means that the sum of the entropies from the
%ordered phase and Schottky-like paramagnetic phase is almost
%constant in magnetic field; the entropy due to the magnetic order
%is transferred to Schottky-like one with the increase of the
%magnetic field, which was reported in \pbnmg\
%already.\cite{02masuda} If we can apply the concept of the
%effective spin to our cases, the magnetic entropy due to effective
%spins should have the same value at the temperature, where its
%entropy is saturated.
%
Hence we can simply assume that the magnetic entropy is the sum of
contribution from the effective spin and spin-gap excitations,
where the former is dominant in low temperature range and the
latter is dominant in high temperature range. In
Fig.~\ref{fig:pbncu-hc} entropies with various \cu\ concentration
seem to converge near 12~K and simultaneously the entropies with
low \cu\ concentration have a point of inflection near the same
temperature. Considering that the spin gap in \pbn\ is about 20 K,
the above experimental facts suggest the saturation of the entropy
due to impurity-induced effective spins and the onset of
contribution from the spin-gap excitation near 12~K. The
experimental value of the entropy at $T \sim$ 12 K is about 0.7 J
Pb-mol$^{-1}$K$^{-1}$ and this is the closest to the saturation
entropy, $Nk_{\rm b}\ln (2 \times \frac{1}{2} + 1) = 0.576$ J Pb-mol$^{-1}$,
calculated based on model (iii) proposed in the section III A.
Hence, we conclude that two edge spins and the spin of impurity
\cu\ ion are {\it antiferromagnetically} coupled and one effective
spin of $S=1/2$ appears near the impurity spin. The conclusion is
consistent with the argument in the section III A.

Magnetic heat capacities and magnetic entropies of \pbnmn\ with
$x=0.010$ and \pbnco\ with $x=0.010$ are shown in
Figs.~\ref{fig:pbnmn-hc} and \ref{fig:pbnco-hc}, respectively. For
Mn-doped sample, an anomaly due to the antiferromagnetic ordering
is found in $H$ = 0 T at 1.1~K, which is the same temperature of
the anomaly in the AC susceptibility (see the inset). The anomaly
disappeared above the magnetic field of $H=2.0$~T. For \pbnco\ a
broad peak of the magnetic heat capacity in zero magnetic field is
observed at $T=2.1$~K, which is the same temperature of the
anomaly in the magnetic susceptibility (see
Fig.~\ref{fig:pbnco-mt} and the inset.) The weak ferromagnetism is
concluded from both heat capacity and the magnetic susceptibility
measurements. The anomaly is suppressed by the application of a
magnetic field and disappears at $H=4.0$~T.

Magnetic heat capacity and entropy in both \pbnmn\ and \pbnco\ are
qualitatively the same as in \pbncu\ and the similar argument
leads to the same conclusion; the model (iii) proposed in the
section III A is the best to describe the effective spin induced
by impurities. $J_M$ in the effective Hamiltonian in Eq.
(\ref{effective_M}) is antiferromagnetic and one effective spin
with $S = |S_M - 1|$ is detected in the magnetic-impurity-doped
\pbn .

\subsection{$T$--$x$ phase diagram\label{IIID}}

\begin{figure}[h!]
    \fig{8}{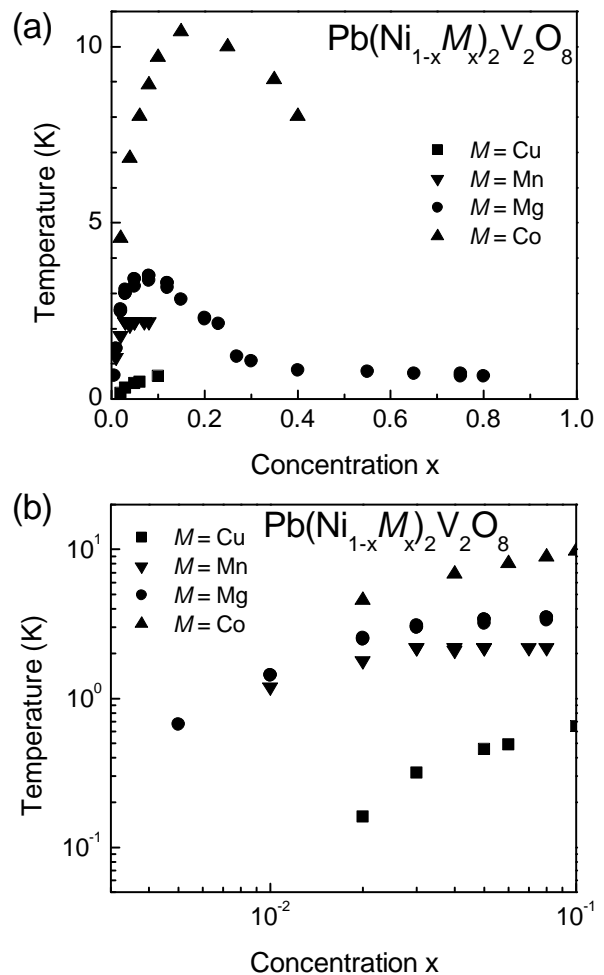}
\caption{(a) Temperature vs impurity concentration ($T$--$x$)
phase diagram in impurity-doped Haldane material \pbn . (b)
$T$--$x$ phase diagram in the lightly doped region, where the
data are plotted in $\log$-$\log$ scales. } \label{fig:tx}
\end{figure}

$T$--$x$ phase diagrams are {\it qualitatively} independent on the
species of impurities as we can see in Fig.~\ref{fig:tx} . The
transition temperatures increase with the impurity concentration
and they reach their maximum at some concentrations. In higher
concentration region the modest decreases are observed on \pbnm\
($M$ = Mg and Co) where the higher concentration samples are
available. We find the similar feature of the $T$--$x$ phase
diagram in other spin-gap materials such as spin-Peierls
\cgo,\cite{96koide,98masuda,00masuda} two-leg spin ladder
SrCu$_2$O$_3$,\cite{97azuma} and interacting dimer
TlCuCl$_3$.\cite{02oosawa} This shows that the phenomenon, i.e.,
impurity-induced antiferromagnetic long-range order in spin-gap
materials, is universal as already discussed briefly in
Refs.~\onlinecite{99uchiyama}, \onlinecite{00uchinokura-a},
\onlinecite{02uchinokura} and \onlinecite{02uchinokura-a}.

We find another common feature in the lightly-doped region. Small
amount of impurities induce antiferromagnetic long-range order and
the transition temperature does not decrease drastically in
log-log scale, which is shown in Fig.~\ref{fig:tx} (b), and this
suggests no threshold concentration in \pbnm . The absence of the
threshold was also reported in Zn-doped \cugeo\ by Manabe
\textit{et~al.}\cite{98manabe} They studied the $T$--$x$ phase
diagram in low-concentration region and showed that the relation
%%%%%%%%%%%%%%%%%%%%%%%%%%%%%%%%%%%%%%
\begin{equation}
T_\mathrm{N}=A\exp\left(-\frac{B}{x}\right),
\end{equation}
%%%%%%%%%%%%%%%%%%%%%%%%%%%%%%%%%%%%%%
holds in very low-temperature ($T\lesssim 8\times 0.8$~K) and
low-concentration ($x\lesssim 5\times 10^{-3}$) region.

%%%%%%%%%%%%%%%%%%%%%%%%%%%%%%%%%%%%%%%%%%%

%%%%%%%%%%%%%%%%%%%%%%%%%%%%%%%%%%%%%%%%%%%

The unique feature in \pbnm\ is that the transition temperatures
drastically depend on the species of the impurities. The maximum
transition temperature of \pbnco\ is more than 10 K, while that of
\pbncu\ is less than 1 K. Non-magnetic impurity Mg$^{2+}$ induces
lower transition temperature than Co$^{2+}$ but, somehow, higher
temperature than Mn$^{2+}$ or Cu$^{2+}$. The seemingly mysterious
behavior can be explained by considering effective Hamiltonian in
Eqs.~(\ref{effective_Mg}) and (\ref{effective_M}). The impurity
dependence of the transition temperatures will be {\it
semiquantitatively} discussed in the following section.

%%%%%%%%%%% DISCUSSION %%%%%%%%%%% DISCUSSION %%%%%%%%%%% DISCUSSION %%%%%%%%%%
\section{Discussion\label{IV}}

The low-temperature phase of \pbnco\ showed a hysteresis in the
susceptibility vs temperature curve and it was not a simple
antiferromagnetic phase. The low-temperature phase of \pbnco\ can
be attributed to the weak-ferromagnetic phase. The reason for the
occurrence of the weak-ferromagnetic phase is due to the structure
of \pbn, where the Ni chain (spin chain) is constructed as the
four-fold screw chain.\cite{86wichmann} The screw chain does not
have an inversion center between the neighboring Ni sites and
therefore the Dzyaloshinskii-Moriya (DM)
interaction\cite{57dzyalo,60moriya} exists between the two
neighboring spins. As is well known, DM interaction can cause
canting of the sublattice magnetization of the antiferromagnetic
phase and the state becomes weak-ferromagnetic phase.

In Fig.~\ref{fig:tx} we see that the transition temperatures of
\pbnm\ differ much depending on the kind of impurities $M$ even
though the qualitative behavior is independent. The transition
temperature is obtained as the divergence of the staggered
susceptibility but there is no theoretical calculation for our
specific case. However, it is possible to {\it semiquantitatively}
explain the $T$--$x$ phase diagram such as extremely low
transition temperature in \pbncu\ and very high transition
temperature in \pbnco .

As was already mentioned, the low energy excitation in the
disordered state in a doped-Haldane chain is well explained by the
effective Hamiltonian in Eq.~(\ref{effective_Hamiltonian}). For
the appearance of the ordered state in coupled-Haldane chains, on
the other hand, the interchain interaction must be considered in
addition. However, the consideration of the effective Hamiltonian
Eq.~(\ref{effective_Hamiltonian}) which includes only $J_{\rm
NNN}$ and $J_M$ would still be meaningful if we assume that
impurity doping does not affect the interchain interaction. We
could discuss the transition temperatures in \pbn\ doped with
various impurities by calculating the ground state energy in
effective Hamiltonian Eqs. (\ref{effective_Mg}) and
({\ref{effective_M}), and also by calculating the energy in local
collinear spin structure; parallel in Fig. \ref{fig:spins} (b) or
antiparallel in Fig. \ref{fig:spins} (c).

Firstly let us examine why \pbncu\ or \pbnmn\ has lower transition
temperatures than \pbnmg . Effective Hamiltonian for \pbnmg\ is
Eq.~(\ref{effective_Mg}) and the ground state is ferromagnetic
triplet state, as is visualized in Fig.~\ref{fig:spins}(b). Such a
collinear state conserves the coherence of antiferromagnetic
correlation in the chain around the impurities. Hence
three-dimensionally ordered state is favored by the local inchain
interaction, ferromagnetic $J_{\rm NNN}$, as well as interchain
interaction. On the other hand the effective Hamiltonian for
\pbncu\ or \pbnmn\ is expressed by Eq.~(\ref{effective_M}). The
local antiparallel spin structure (see Fig.~\ref{fig:spins} (c)),
which favors coherence of antiferromagnetic correlation in the
chain, is neither the ground state nor an eigenstate of the
effective Hamiltonian. This means that the three-dimensionally
ordered state in \pbncu\ or \pbnmn\ is favored by interchain
interaction but it is disturbed by the local inchain interaction,
$J_M$. Therefore \pbncu\ or \pbnmn\ has lower transition
temperatures than \pbnmg .

\begin{table*}
\caption{\label{tab:table1}Summary of the maximum transition
temperature, the ground state energy, collinear spin state energy,
and the scaled energy difference. Details are explained in the
text. }
\begin{ruledtabular}
\begin{tabular}{lcccc}
&$M$ = Cu&$M$ = Mn &$M$ = Mg &$M$ = Co\\ \hline $T_{\rm N}^{\rm
max}$ & 0.65 K ($x$ = 0.1) & 2.2 K ($x$ = 0.05) & 3.5 K ($x$ =
0.08) & 10.4 K ($x$ = 0.15)\\ $\mathcal{H}$ & $\mathcal{H}_2$ &
$\mathcal{H}_2$ & $\mathcal{H}_1$ & $\mathcal{H}_3$ \\$E_{{\rm
G},M}$ & $-J_{\rm Cu}+1/4J_{\rm NNN}$ & $-7/2J_{\rm Mn}+1/4J_{\rm
NNN}$ &
1/4$J_{\rm NNN}$ & $-3/2J_{\rm Co} + 1/4J_{\rm NNN}$\\
$E_{{\rm collinear},M}$ & $-1/2J_{\rm Cu}+1/4J_{\rm NNN}$ &
$-5/2J_{\rm Mn}+1/4J_{\rm NNN}$ &
1/4$J_{\rm NNN}$ & $-3/2J_{\rm Co} + 1/4J_{\rm NNN}$\\
$\Delta E_M$ & 2 & 4/5 & 0 & 0\\
\end{tabular}
\end{ruledtabular}
\end{table*}

Secondly let us discuss the transition temperatures in \pbncu\ and
\pbnmn . The ground states of both of the effective Hamiltonians
are not a collinear spin state. This fact does not favor the
antiferromagnetic coherence in the isolated chain and the
appearance of ordered state is the result of the reduction in the
energy of the ordered state due to the interchain interaction. In
this case the transition temperature may be mainly determined by
the energy difference between the ground state of the effective
Hamiltonian and collinear spin state. We define
\begin{equation}
\Delta E_{M = {\rm Cu, Mn}} \equiv \Bigl(\frac{E_{\rm
collinear}-E_{\rm G}}{J_Ms_1S_M}\Bigr)_{M = {\rm Cu, Mn}},
\label{energydifference}
\end{equation}
where
\begin{eqnarray}
E_{{\rm collinear},M} &=& \langle {{\rm
collinear},M}|\mathcal{H}_2|{{\rm collinear},M}
\rangle \nonumber \\
|{{\rm collinear},M} \rangle &=& | s_1^z=1/2, S_M^z=-S_M,
s_2^z=1/2 \rangle .
\end{eqnarray}
For the appearance of three-dimensionally ordered state, large
$\Delta E_{M}$ requires large energy reduction in the ordered
state and this results in the lower transition temperature. The
maximum transition temperatures $T_{\rm N}^{\rm max}$, $E_{{\rm
G},M}$, $E_{{\rm collinear},M}$, and $\Delta E_M$ are summarized
in Table~\ref{tab:table1}. We find $\Delta E_{\rm Cu}
>\Delta E_{\rm Mn}$ and this is consistent with lower transition
temperature in \pbncu\ than in \pbnmn .

Thirdly let us discuss the highest transition temperatures in
\pbnco . Since we have already explained that \pbnmg\ has the
highest transition temperature among Mg-, Cu-, and Mn-doped \pbn ,
the comparison between \pbnco\ and \pbnmg\ is enough. So far we
assumed implicitly that the spin interactions are isotropic. In
some magnetic ions such as Co$^{2+}$, however, the interaction is
possibly Ising-like because of remnant orbital momentum. In this
case the first and the second term in Eq.~(\ref{effective_M}) are
modified;
\begin{equation}
\mathcal{H}_3=J_{\rm Co}(s^z_1\cdot S^z_{\rm Co}+S^z_{\rm Co}\cdot
s^z_2)+J_{\rm NNN}\bm{s}_1\cdot\bm{s}_2. \label{effective_Ising}
\end{equation}
The ground state of the effective Hamiltonian is spin collinear
structure, as is similar to \pbnmg , and the lower the ground
state energy is, the higher the transition temperature will be.
Useful physical quantities for \pbnmg\ and \pbnco\ are summarized
in Table~\ref{tab:table1}. The ground state energy in \pbnco\ is
lower than that of \pbnmg\ by $-(3/2)J_{\rm Co}$ and, therefore,
the former
 has higher transition temperature.
This means that \pbnco\ has the highest transition temperature.

The $T$--$x$ phase diagram in \pbnm\ has been {\it
semiquantitatively} explained by the simple effective Hamiltonians
based on VBS model. The drastic difference between magnetic
impurity (Cu$^{2+}$ or Mn$^{2+}$) and non-magnetic impurity
(Mg$^{2+}$) is due to the different ground states. The highest
transition temperature in \pbnco\ could be ascribed to Ising-type
interaction. The effective Hamiltonians were successfully applied
to Haldane materials but they can not be applied to other spin-gap
materials such as $S=1/2$ spin-Peierls \cugeo. Indeed there is no
drastic difference in the transition temperatures between
non-magnetic impurity (Mg$^{2+}$ or Zn$^{2+}$) and magnetic
(Ni$^{2+}$ $S$ = 1) impurity-doped \cugeo .\cite{02smasuda}

In doped \cugeo\ some of the present authors found that there is a
first-order phase transition between the two kinds of
antiferromagnetic phases with the change of the concentration of
non-magnetic impurity \mg\ (Refs.~{\onlinecite{98masuda} and
\onlinecite{00masuda}}) or \zn.\cite{00masuda} Two phases are the
dimerized-antiferromagnetic (DAF) phase and
uniform-antiferromagnetic (UAF) phase.\cite{98masuda} This
phenomenon has been studied in detail by the various methods;
thermal conductivity,\cite{00takeya} Raman
scattering,\cite{99kuroe} neutron
diffraction,\cite{99nakao-a,99nakao-b,00nishi} pressure
effect,\cite{01masuda,03tanokura,03masuda} ESR,\cite{02glazkov}
and synchrotron x-ray scattering.\cite{03wang} In
Fig.~\ref{fig:tx}(a) we see that the antiferromagnetic phase of
\pbnmg\ extends very widely to $x\sim 1$ (PbMg${}_2$V${}_2$O${}_8$
is not magnetic) and there seems no compositional phase transition
in this wide region. On the contrary in Mg-doped \cugeo\ the
concentration region of the antiferromagnetic phase is more
limited and there is a first-order compositional phase transition
as mentioned above. The concept of the ``impurity-induced
antiferromagnetic phase" may only be applied to the lightly doped
region in \pbnmg. When the concentration $x$ of Mg  is very large,
magnetic \nic\ ions are randomly distributed three dimensionally
among the non-magnetic \mg\ ions and the ordered state may be
constructed among these diluted \nic\ spins. In \cugeo\ the
spin-gap state is caused by the spin-lattice interaction. In
\cugeo\ and doped \cugeo\ spin gap is created by the spin-Peierls
transition and the antiferromagnetic phase in the low impurity
concentration region (DAF phase) has two order parameters:
dimerization and antiferromagnetic long-range order. The
difference of DAF and UAF phases in doped \cugeo\ is the existence
or absence of the lattice distortion (dimerization). On the
contrary in \pbn\ the spin gap is intrinsic and in the
impurity-induced antiferromagnetic phase there is ony one order
parameter (antiferromagnetic long-range order). The difference of
the numbers of the order parameters may cause the different
behavior of the compositional change between
Cu${}_{1-x}$Mg${}_x$GeO${}_3$ and \pbnmg.

%%%%% SUMMARY %%%%%%%%%%%%%%% SUMMARY %%%%%%%%%%% SUMMARY %%%%%%%%%%%%%%%
\section{Summary}
Impurity-induced three-dimensional ordering is systematically
studied in doped Haldane material \pbn\ by use of DC and AC
magnetic susceptibility, and heat capacity, on many samples with
various species of impurities and concentrations. Complete
$T$--$x$ phase diagrams are obtained for each impurity and
qualitatively common features to the doped spin-gap materials are
observed. The unique feature is found in the drastic dependence of
the transition temperatures on the species of the impurity, which
is explained by effective Hamiltonian based on VBS model.

%%%%%%Acknowledgements%%%%%%%%%%%%%%%%%%%%%%%%
\begin{acknowledgements}
This work was partially supported by Grant-in-Aid for COE Research
``SCP Project" from the Ministry of Education, Culture, Sports,
Science, and Technology of Japan. One of the authors (T.~M.) was
partially supported by DOE Contract No. DE-AC05-00OR22725.
\end{acknowledgements}

%%%%%%%%%% REFERENCE %%%%%%%%%%%%% REFERENCE %%%%%%%%%%%% REFERENCE %%%%%%%%%%%

%\bibliography{pbn-new_3}

\end{document}